\documentclass[aps, showpacs, twocolumn, amssymb, superscriptaddress]{revtex4-1}

\usepackage{graphicx}

\begin{document}

\title{Correction to the detector-dead-time effect on the second-order correlation of stationary sub-Poissonian light in a two-detector configuration}

\author{Byoung-moo Ann}
\affiliation{Department of Physics and Astronomy, Seoul National University, Seoul 151-747, Korea} 
\author{Younghoon Song}
\author{Junki Kim}
\author{Daeho Yang}
\author{Kyungwon An}
\email{kwan@phya.snu.ac.kr}
\affiliation{Department of Physics and Astronomy, Seoul National University, Seoul 151-747, Korea} 
\date{\today}

\begin{abstract}
Exact measurement of the second-order correlation function $g^{(2)}(t)$ of a light source is essential when investigating the photon statistics and the light generation process of the source.
For a stationary single-mode light source, Mandel $Q$ factor is directly related to $g^{(2)}(0)$. 
For a large mean photon number in the mode, the deviation of $g^{(2)}(0)$ from unity is so small that 
even a tiny error in measuring $g^{(2)}(0)$ would result in an inaccurate Mandel $Q$.
In this work, we have found that detector dead time can induce a serious error in $g^{(2)}(0)$ 
and thus in Mandel $Q$ in those cases even in a two-detector configuration.
Our finding contradicts the conventional understanding that detector dead time would not affect $g^{(2)}(t)$ in two-detector configurations.
Utilizing the cavity-QED microlaser, a well-established sub-Poissonian light source, we measured $g^{(2)}(0)$ with two different types of photodetectors with different dead time. We also introduced prolonged dead time by intentionally deleting the photodetection events following a preceding one within a specified time interval.
We found that the observed $Q$ of the cavity-QED microlaser was underestimated by 19\% with respect to the dead-time-free $Q$ when its mean photon number was about 600.
We derived an analytic formula which well explains the behavior of the $g^{(2)}(0)$ as a function of the dead time. 
\end{abstract}
\preprint{Prepared for PRA}

\maketitle

\section{Introduction}\label{sec1}
The second-order correlation (SOC) function $g^{(2)}(t)$ of radiation is a key quantity characterizing photon statistics as well as elucidating the underlying light generation mechanism.  
This correlation function is often interpreted as being proportional to the probability of measuring a photon at time zero and then measuring another photon at time $t$.
Since the first observation of SOC of light from a star by Hanbury Brown and Twiss (HBT) \cite{Hanbury-Brown-57}, the measurement techniques for SOC have progressed much as high-efficiency photo detectors and fast electronics have been developed. 

Significant improvements were made by employing a time-to-digital converter or a time digitizer, which provides a digital representation of the time intervals between a start photodetection event at one detector and multiple stop events at the other detector. 
From these time intervals a histogram of the time delay between the start and stop events is obtained.
This method is called the single-start multi-stop time-to-digital conversion(SMTDC) \cite{Cummins-77,Swinney-83,Pike-83}. 
In later experiments, a more efficient method which uses all of the arrival time records in both start and stop detectors was developed. This method is called the multi-start multi-stop time-to-digital conversion (MMTDC) compared to the conventional SMTDC. 
In MMTDC, all of the arrival times at both detectors are recorded for a time window $T_0$ much longer than the correlation time $\tau_c$ of a radiation source. 
By software or by using a hardware correlator, we then obtain the correlation of all detected photon pairs or more specifically 
a histogram of time intervals between all possible detected photon pairs.
The number of detection events during $T_0$ on a start detector in MMTDC is given by $N_0=\eta\Phi T_0$, where 
$\eta$ is the quantum efficiency of the detector and $\Phi$ is the incident photon flux.
Therefore, MMTDC is more efficient than SMTDC, which uses only one start photon event, by this factor of $N_0\gg 1$.
Owing to this high efficiency, MMTDC has been successfully employed in the first observation of non-classical radiation \cite{Choi-PRL06} and quantum frequency pulling \cite{Hong-PRL12} in the cavity-QED microlaser and the spectrum of a single atom localized in an optical lattice \cite{Kim-NL12}.

For accurate measurement of SOC, the effects of detector characteristics such as detection efficiency and dead time have also been investigated, where the latter is a time period in which a photo detector becomes blind after photodetection. 
Although the SOC function $g^{(2)}(t)$ of light is independent of detector efficiency, 
it is apparent that in a single detector configuration
detector dead time $\tau_d$ can seriously affect the measurement of $g^{(2)}(t)$. 
Because of the detector dead time,
$g^{(2)}(t)$ is significantly reduced for $|t|<\tau_d$ near the origin. As a result, we lose the information on $g^{(2)}(0)$, an important parameter directly related to the photon statistics of a stationary radiation source as discussed below. 
No effective way to recover the lost information completely has been found for a single detector configuration despite many studies on this issue \cite{Apansovich-JOSA09,Larsen-MST09}.

It is often argued that the dead-time deficiency may be completely removed in a two-detector configuration for stationary light sources. 
This is based on a simple reasoning that two successive photons within the detector dead time can be resolved
if those two photons are detected on separate detectors: the first photon is detected on a start detector and the second photon is on a separate stop detector.
For non-stationary light sources, on the other hand, detector dead time has been shown to distort SOC functions as discussed by Choi {\em et al.} \cite{Choi-RSI05} even in a two-detector configuration. 

In this paper, we show that the detector-dead-time deficiency cannot be completely removed in the two-detector configuration even for stationary light sources. In this case, dead time effect comes in as reduction in the detected flux due to the missed photons during $\tau_d$ compared to the mean waiting time $\tau_w$ -- the mean time interval between successive photodetection events.
As a result, non-negligible distortion occurs in the SOC function $g^{(2)}(t)$ for $|t|<\tau_d$. 
The distortion deepens as the incident photon flux $\Phi$ increases and consequently the mean waiting time $\tau_w=1/(\eta\Phi)$ is reduced to approach $\tau_d$.
Such distortion is critical
especially for a nonclassical light source with a large internal mean photon number $\langle n\rangle$. 
For a stationary single mode light, the relation $g^{(2)}(0)=1+Q/\langle n\rangle$ \cite{Fox-2006} holds with the Mandel $Q$ parameter bounded between -1 and 0 for nonclassical light. For a large mean photon number $\langle n\rangle \gg 1$, we then have $|1-g^{(2)}(0)|\ll 1$, so even small distortion by the detector dead time can cause a large error in determining $Q$. 
In this work, we first derive a formula quantifying the distortion in $g^{(2)}(0)$ induced by the dead time and then verify its validity in actual experiments with the cavity-QED microlaser \cite{Choi-PRL06} generating a stationary nonclassical radiation. We show that by using the formula we can recover $g^{(2)}(0)$ and thus the Mandel $Q$ unaffected by the detector dead time.

This paper is organized as follows. 
In Sec.\ \ref{sec2}, we discuss dead time effect on mean photon flux measurement. 
We then extend our discussion to SOC measurement and derive a formula to correct the distortion introduced by the dead time in $g^{(2)}(0)$ in Sec.\ \ref{sec3}.
Our experimental setup and simulation methods for checking the validity of our formula are discussed in Sec.\ \ref{sec4}. 
We present the experimental and simulation results consistent with our theoretical description in Sec.\ \ref{sec5} followed by concluding remarks in Sec.\ \ref{sec6}.

\section{Dead time effect on photodetection flux}\label{sec2}
Two-detector configuration eliminates the distortion due to the missed successive photons on the same detector by considering two successive photon counting events only on separate start and stop detectors. However, there still exists another source of distortion coming from the reduction in counted photon flux due to the dead time.

\subsection{light with Poisson photon statistics}
Let us first consider a waiting-time distribution $w(t)$ for a detector with a quantum efficiency $\eta$ but without dead time.
If the photon statistics of light is Poissonian like coherent light, the waiting-time distribution is given by a single exponential function: $w(t)=\phi e^{-\phi t}$ with $\phi=\eta\Phi$, the dead-time-free photodetection flux for the incident light.
The mean waiting time $\tau_w$ is then given by $\tau_w=\phi^{-1}$. In the presence of detector dead time, the waiting-time distribution is modified in such a way that it vanishes for $0<t<\tau_d$ with the rest still the same exponential as $w(t)$. When normalized, the modified waiting-time distribution $w'(t)$ is nothing but $w(t-\tau_d)$. 
It is then straightforward to show that the new mean waiting time $\tau'_w$ is given by 
\begin{equation}
\tau'_w=\tau_d+\tau_w
\label{eq0}
\end{equation}
In the presence of the detector dead time, the photodetection flux $\phi'=1/{\tau'_w}$ for the incident light appears to be less than the dead-time-free photodetection flux $\phi$ by the following relation. 
\begin{equation}
\phi'=\frac{1}{\tau_w+\tau_d}=\frac{\phi}{1+\phi\tau_d}
\label{eq1}
\end{equation}
This formula is already derived in the previous works \cite{Albert-NMS53, Takacs-AMS58}. It has been used to investigate the dead time effect on the intensity statistics of scattered light field measured with a finite collection aperture \cite{Schztzel-APB86, Schatzel-JOSAB89}.
We can then interpret the quantity $\phi' / \phi$ as ``capture probability'' and $1-(\phi' / \phi)$ as ``miss probability'' in photodetection due to the dead time, respectively.

\subsection{light with non-Poissonian statistics}\label{sec2B}
If the light source exhibits sub- or super-Poisson photon statistics, the waiting time distribution is not given by a simple exponential function and thus Eq.\ (\ref{eq1}) is no longer valid in general.
For instance, let us consider light exhibiting sub-Poisson photon statistics with its SOC function given by 
$g^{(2)}(t)=1- e^{-t/\tau_c}$ with $\tau_c$ the correlation time. 
Ververk and Orrit \cite{Ververk-JCP03} showed that 
the waiting-time distribution is approximately double-exponential, given by $w_0(t)\simeq\phi (e^{-\phi t}-e^{-t/\tau_c})$ for an ideal detector of $\eta=1$. The relation between $\phi'$ and $\phi$ would then be quite different from Eq.\ (\ref{eq1}). 

In general, the capture probability can be written in terms of the detector dead time as
\begin{equation}
\left(\frac{\phi'}{\phi}\right)^{-1}=1+\sum_{n=1}^\infty a_n x^n 
\label{eq1'}
\end{equation}
where $x=\phi \tau_d$ and the coefficient  $a_n$ is given by
\begin{equation}
a_n=\frac{1}{n!}\left.\frac{d^n (\phi/\phi')}{dx^n} \right|_{x=0}
\label{eq-an}
\end{equation}
depending on the specific waiting time distribution of the system under consideration.
For the above waiting time distribution $w_0(t)$, the lowest non-vanishing coefficient is $a_2=[2(1-\phi \tau_c)\phi \tau_c]^{-1}$ under the condition $\tau_d < \tau_c$.

If the mean internal photon number $\langle n\rangle$ of a source is much larger than $|Q|$ regardless of its photon statistics, the SOC function is close to that of coherent light, {\em i.e.},  $|1- g^{(2)}(t)|\ll 1$, and the corresponding waiting-time distribution is approximately single-exponential. An example is 
the cavity-QED microlaser, where $\langle n\rangle \sim 10^2 - 10^3$ and $-1<Q<1$. We can then use Eq.~(\ref{eq1}) for consideration of a detector dead time on the photodetection flux. The precise condition for the validity of this approximation is $|1-g^{(2)}(0)|\ll 1$  as shown in Appendix.

Ververk and Orrit neglected the higher-order correlation effects in deriving the above waiting-time distribution. More general discussion including the higher-order correlations is presented in Ref.~\cite{Barsegov-JCP02, Barsegov-JCP02b}. We also neglect the higher-order correlations $g^{(n)}(0,\ldots, 0,t)$ with $n\ge 3$ in this paper. It is because that the higher-order correlations are not much different from SOC, unlike thermal light sources \cite{Hong-Thesis11}, in the cavity-QED microlaser, where the condition $|1-g^{(2)}(0)|\ll 1$ is well satisfied.

\section{Dead time effect on second-order correlation measurement}\label{sec3}

In a two-detector configuration, the intensity correlation $\langle I_{st}(t)I_{sp}(t+t')\rangle$ is understood as a joint probability of photodetection at time $t$ on a start detector and at time 
$t+t'$ on a stop detector. The intensity operator for the start(stop) port is denoted as $I_{st}(I_{sp})$.
For stationary light, the correlation does not depend on $t$ and thus it can be replaced with a fixed time.
If we define $\mathcal{N}(t)$ to be the actual number of photon pairs comprised of one photon incident on the start detector and another on the stop detector with a time delay $t$, the normalized SOC function $g^{(2)}(t)$ in this configuration can be expressed as 
\begin{equation}
g^{(2)}(t)=\frac{\langle I_{st}(0)I_{sp}(t)\rangle}{\langle I_{st}(0)\rangle \langle I_{sp}(t)\rangle}=\frac{\mathcal{N}(t)}{\mathcal{N}(\infty)}=\frac{N(t)}{N(\infty)},
\label{eq2}
\end{equation}  
where $N(t)$ is the number of photodetection pairs with a time delay $t$ in the absence of detector dead time, so $N(t)=\eta^2 \mathcal{N}(t)$.
Equation~(\ref{eq2}) shows that SOC does not depend on the detector efficiency.

\begin{figure}
\includegraphics[width=3.4in]{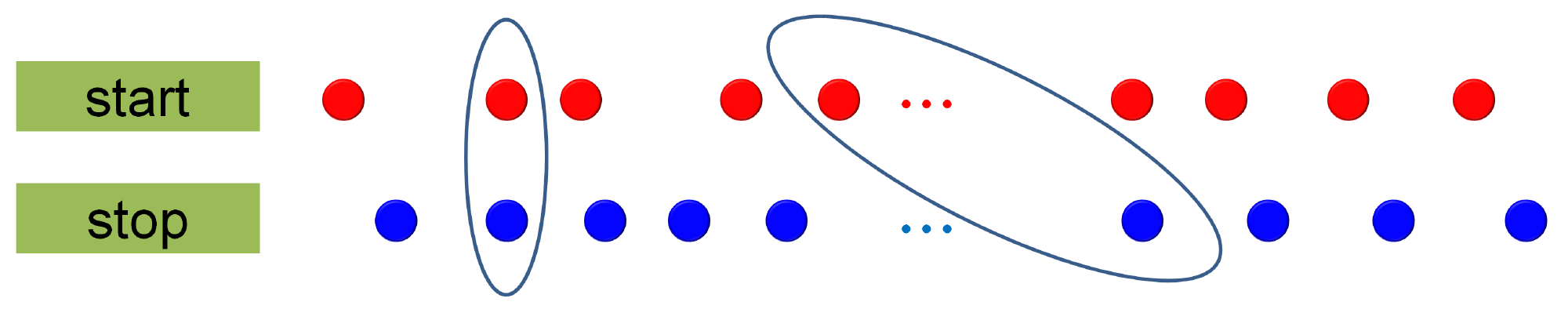}
\caption{
Illustration of photodetection events in a HBT-type two-detector configuration. The number of photon pairs with a zero time delay (circled on the left) and that with a very long time delay (circled on the right) are needed in evaluating $g^{(2)}(0)$.}
\label{fig1}
\end{figure}

From now on, let us concentrate on $g^{(2)}(0)$, a parameter directly related to the Mandel $Q$ of the internal field of the source.
In evaluating $g^{(2)}(0)$ with Eq.~(\ref{eq2}), the numerator $\mathcal{N}(0)$ is obtained by counting events like the circled one on the left in Fig.\ \ref{fig1} whereas the denominator $\mathcal{N}(\infty)$ is obtained from the events like the one circled on the right.
Our interest is then how the detector dead time affects such counting events.
We restrict ourselves to the case of $|1- g^{(2)}(t)|\ll 1$, {\em i.e.}, to the case where the waiting-time distribution is near single-exponential and thus we can still use  Eq.\ (\ref{eq1}).

Let us first consider the time delay $t$ much larger than a correlation time. In this case, we can neglect the correlation between photons in each pair. 
Because of the detector dead time, each photo counting event is then less probable by the capture probability $\phi'/\phi=(1+\phi\tau_d)^{-1}$, so the denominator $\mathcal{N}(\infty)$ should be replaced by
\begin{equation}
\mathcal{N}(\infty) \rightarrow \frac{\mathcal{N}(\infty)}{\left(1+\phi_{st}\tau_d\right) \left(1+ \phi_{sp}\tau_d\right)},
\label{eq3}
\end{equation}
where $\phi_{st} (\phi_{sp})$ is the dead-time-free photodetection flux on the start(stop) detector.

For zero time delay, on the other hand, the photo flux on each detector is further modified by a factor $g^{(2)}(0)$. It is because the probability of having a photon on one detector with another photon on the other detector at the same time is proportional to $g^{(2)}(0)$. Including this effect, the photo detection flux $\phi$ is replaced by $g^{(2)}(0)\phi$ for each detector. The numerator $\mathcal{N}(0)$ should then be replaced by
\begin{equation}
\mathcal{N}(0) \rightarrow \frac{\mathcal{N}(0)}{\left(1+g^{(2)}(0)\phi_{st}\tau_d\right) \left(1+ g^{(2)}(0)\phi_{sp} \tau_d\right)}.
\label{eq5}
\end{equation}
As a result, the observed SOC $g'^{(2)}(0)$ will be 
\begin{equation}
g'^{(2)}(0) =  g^{(2)}(0) 
\frac{1+ \phi_{st} \tau_d}{1+g^{(2)}(0)\phi_{st} \tau_d}
\frac{1+ \phi_{sp} \tau_d}{1+g^{(2)}(0)\phi_{sp} \tau_d}.
\label{eq6}
\end{equation}

When the photodetection flux on each detector is as low as $\phi\tau_d\ll 1$, Eq.~(\ref{eq6}) is further approximated as
\begin{equation}
g'^{(2)}(0) \simeq {g^{(2)}(0)} \left\{ 1+\left [1-g^{(2)}(0) \right ](\phi_{st}+\phi_{sp}) \tau_d\right\},
\label{eq7}
\end{equation}
or in terms of Mandel $Q$
\begin{equation}
Q'\simeq  Q \left[1-g^{(2)}(0) (\phi_{st}+\phi_{sp})\tau_d \right].
\label{eq7'}
\end{equation}
This approximation shows that the dead time effect becomes important as the photodetection flux increases for a given dead time.

For an arbitrary waiting time distribution, the capture probability of photodetection under detector dead time can be written as that in Eq.\ (\ref{eq1'}). If the lowest non-vanishing term in the denominator is of the $n$-th order, Eq.\ (\ref{eq7'}) is then replaced with
\begin{equation}
 Q' \simeq Q \left\{ 1-a_n g^{(2)}(0) (\phi_{st}^n+\phi_{sp}^n)\tau_d^n\right\} .
\label{eq7-2}
\end{equation}
with $a_n$ is given by Eq.\ (\ref{eq-an}).

\section{Experiment and Dead-Time Simulation}\label{sec4}
\subsection{Counter Electronics}   
The counter electronics used in the present experiment is an improved version of the one used by Choi {\em et al.} \cite{Choi-PRL06,Choi-RSI05}. 
In the former system, two counter/timing boards were separately installed in two personal computers (PC's) in order to avoid interchannel crosstalk. Another PC was used to trigger those counter/timing boards and to control overall measurement sequence.
In this configuration, each counter/timing board, once triggered, records photodetection times based on its own internal clock, and thus the clocks in those counting channels were not synchronized.

Clock synchronization is realized in the present setup by employing a counter board (National Instruments NI-7813R) equipped with a field programmable gate array (FPGA). 
By programming the FPGA we have implemented multiple counting channels without crosstalk in a single board. 
Those counting channels are perfectly synchronized at a clock speed of 125 MHz while internally triggered. The resulting time resolution is 8 ns, improved from 12.5 ns of the former system.

\begin{figure}
\includegraphics[width=3.4in]{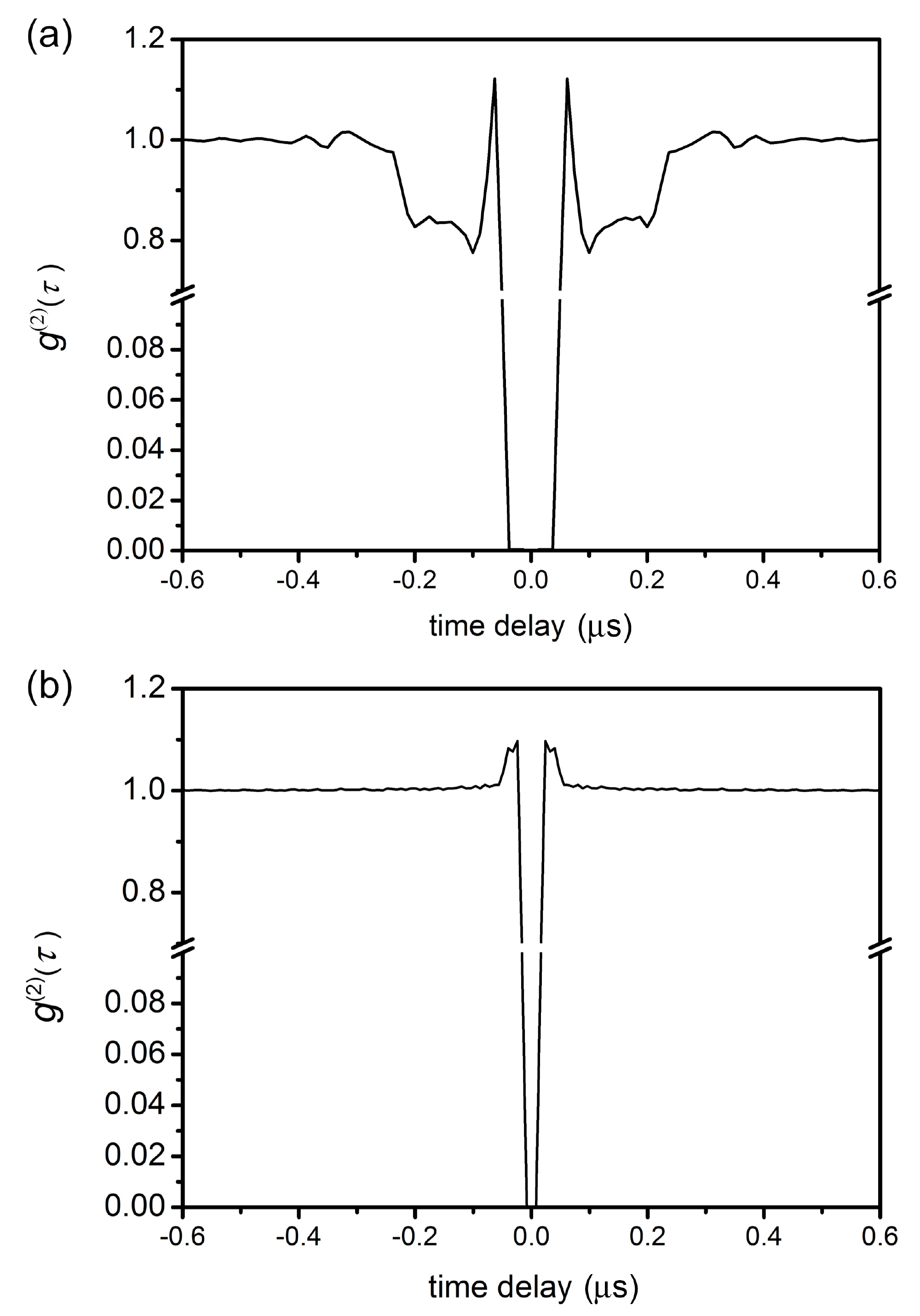}
\caption{Counter-board-related dead-time effect in the observed $g^{(2)}(t)$ in a single detector configuration. (a) With the previous counter/timing board with an onboard register. (b) With the new FPGA counter board with an onboard FIFO memory. Poissonian background light was used as a light source. The partial dips in the range of 100 ns $<|t|<$ 250 ns are due to the data transfer loss between the counter/timing board and a control computer.}
\label{fig2}
\end{figure}

Moreover, with the present setup using an FPGA we have eliminated the counter-board-related dead-time effect observed in the previous setup.
When a photodetection event occurs, the counter board in the previous setup saves the event time measured in clock period in an onboard register first and then it is transferred to the computer memory through direct memory access (DMA). 
A problem arises when the next photodetection event occurs before this transfer is completed: the new event is simply ignored. As a result, the SOC  in a single channel configuration exhibits a partial dead-time effect as shown in Fig.\ \ref{fig2}(a), 
where the SOC function obtained for coherent light is plotted. 
The narrow perfect dip in the range of $|t|<50$ ns is due to the detector dead whereas
the wide partial dips in the range of 100 ns $<|t|<$ 250 ns arises from the above data loss at the counter board. 
We call this the counter dead-time effect. 
In the present setup using an FPGA, the onboard first-in-first-out (FIFO) memory acts as a large buffer in the DMA transfer and thus can eliminate the above counter dead-time effect.
Its performance in a single-channel SOC function measurement is shown in Fig.\ \ref{fig2}(b), where only the detector dead time effect is noticed in the range of $|t|<$ 20 ns from without any partial dips due to the counter dead time.
Clear isolation of the detector dead time effect as in Fig.\ \ref{fig2}(b) in fact enables us to correct the SOC function against the detector dead time in Sec.\ \ref{sec5}.

\begin{figure}
\includegraphics[width=3.4in]{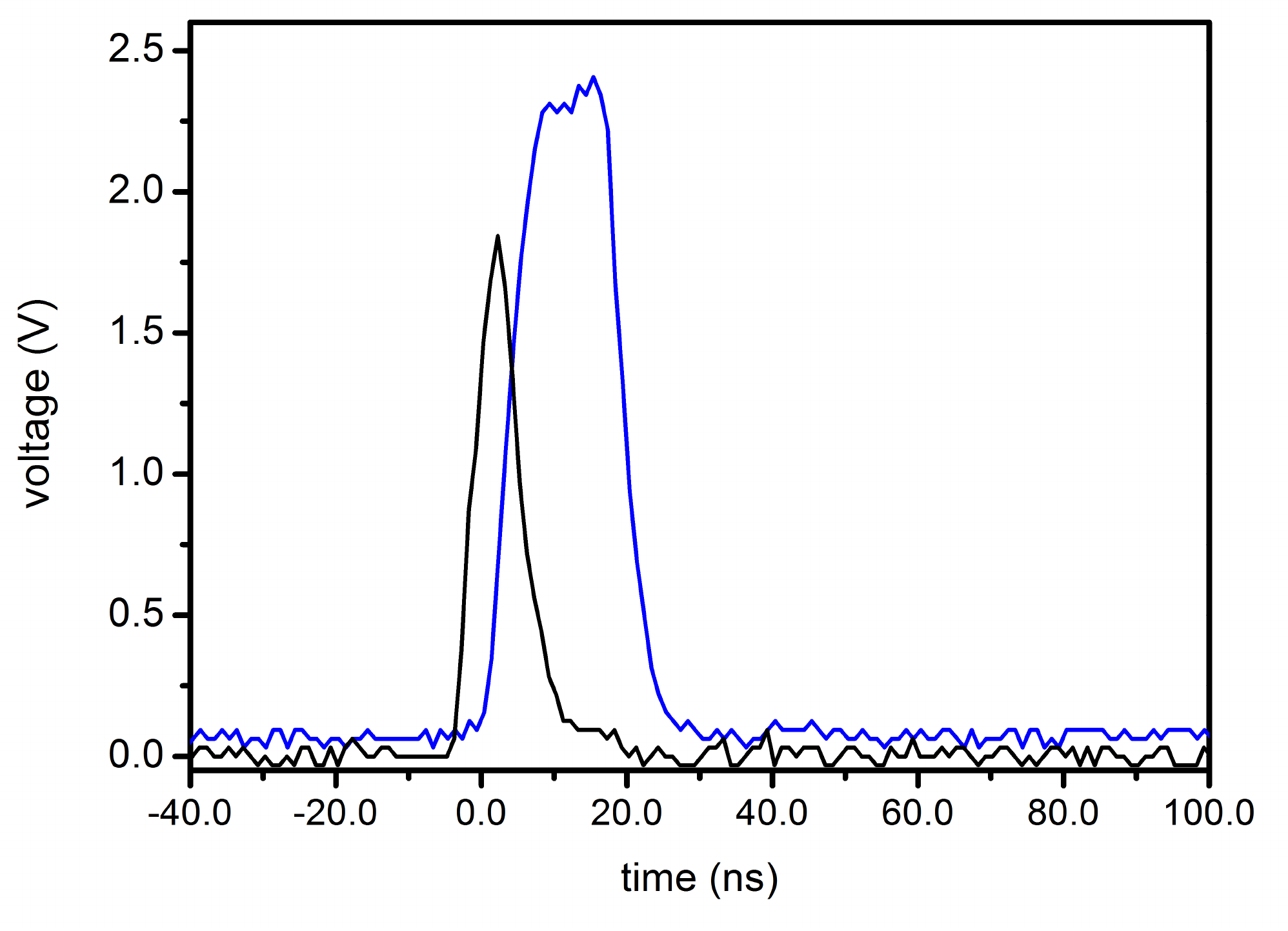}
\caption{
The shape of a pulse (black) from a SPCM-F and that of a pulse (blue) processed by a homemade pulse stretcher. In the pulse stretcher, the original pulse and its delayed pulse are added in time with a small gain by using a OR gate. A delay of a few nanoseconds is due to the intrinsic time delay of the gate chip.
}
\label{fig3}
\end{figure}
\begin{figure}

\includegraphics[width=3.4in]{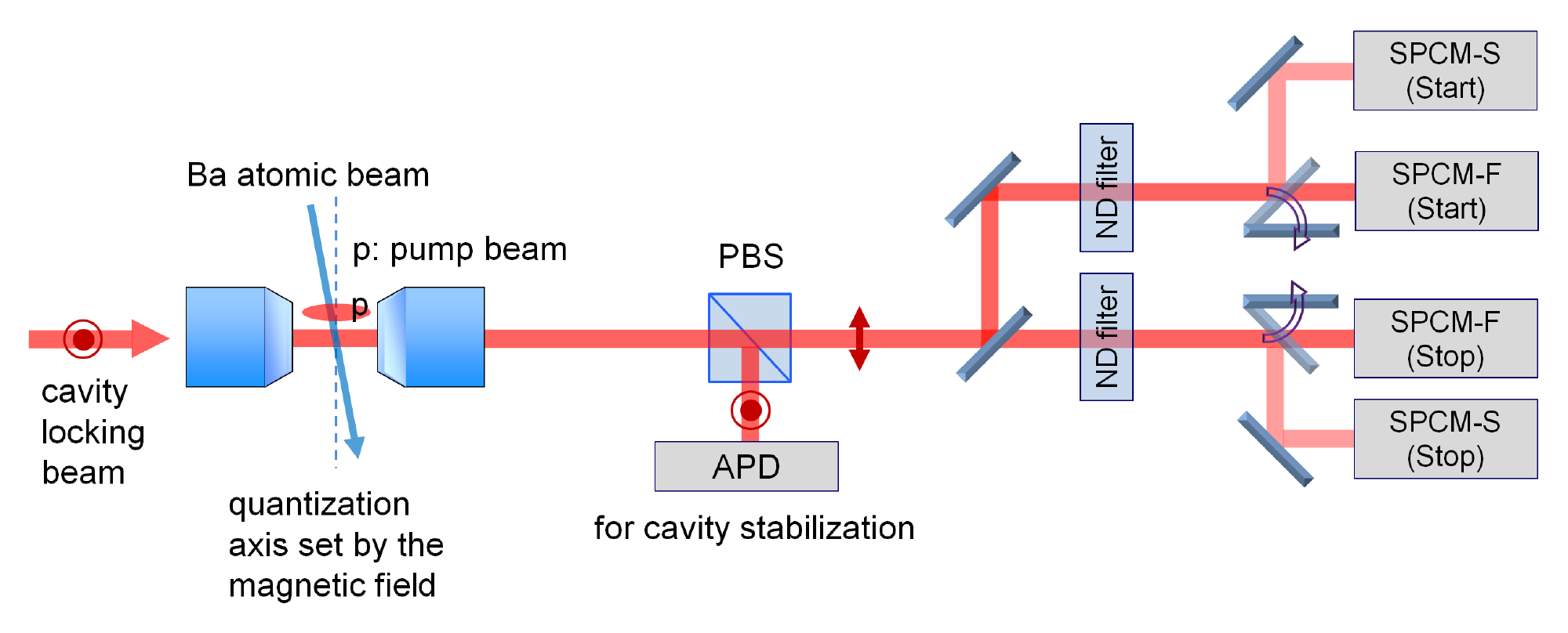}
\caption{
Experimental schematic.
The cavity-QED microlaser is pumped by a beam of barium atoms prepared in the excited state \cite{Choi-PRL06}. The SOC of the output is measured in a two-detector configuration. 
Flippable mirrors (FM's) are used to select a desired SPCM pair, SPCM-F or SPCM-S, while keep the other experimental conditions unchanged.
}
\label{fig4}
\end{figure}

\subsection{Single-Photon Counting Detectors}

Two different models of single-photon counting modules (SPCM's) are used in our experiment. One has a dead time of 50 ns (Perkin Elmer SPCM-AQR-12) and it will be referred as SPCM-S (slow). 
The other has a shorter dead time of 21 ns (Excelitas SPCM-AQRH-12) and it will be referred as SPCM-F (fast).
They have the same characteristics except for the dead time and output voltage specification. 
The output voltage pulse of SPCM-F is not compatible with our counter board (NI-7813R). A homemade pulse stretcher made of fast logic gates is used between them: the output pulse width of 7 ns of SPCM-F is extended to 12.5 ns with an enhanced peak height for the counter board as shown in Fig.\ \ref{fig3}. 

\subsection{Experiment}
Our experimental schematic is depicted in Fig.\ \ref{fig4}. The basic physical principles and apparatus to generate sub-Poisson light with the cavity-QED microlaser is the same as in the previous work by Choi {\em et al.} \cite{Choi-PRL06}. 
In order to facilitate the switching between the two types of detectors with different dead times,
flippable mirrors are used to provide a choice of detectors while preserving the other experimental conditions. 

The SPCM manufacturers provide empirical counting correction factors in the instruction manual \cite{Perkin-Elmer05} up to photodetection flux of  $2.5\times 10^7$ counts/sec or 25 Mcps (Mega counts per second).  These correction factors start to deviate from that in Eq.~(\ref{eq1}) at photon flux of 4 Mcps 
by unexplained reasons. In order to avoid this inconsistency at high photon flux, we attenuate the photon flux to keep the photodetection flux under 3 Mps, where the manufacturers' correction factors well agree with Eq.~(\ref{eq1}).

We have also measured the detector dead times from the actual waiting-time distributions. With a combination of two SPCM-F's  and the FPGA counting board, the dead time extracted from the waiting-time distribution was 28 ns. For SPCM-S, the observed dead time was 56 ns. Detection bin time was 8 ns for both cases.  

We neglect the effect of after-pulsing in our measurement because the probability of after-pulsing is only 0.3\% per real photodetection according to the detector manual. 
Furthermore, after-pulses at separate detectors are perfectly uncorrelated, and thus they contribute a Poissonian background in SOC measurement. Another Poissonian background associated with the detector dark counts is less than 0.1\% of the actual photon flux in our experiment, so it does not affect our correlation measurement either. 
If a Poissonian background count rate were comparable to the actual photon flux, detector dead time could make a huge difference in measuring SOC in a different way as previously studied in Ref.~\cite{Zhang-PRA05}.

\begin{figure}
\includegraphics[width=3.4in]{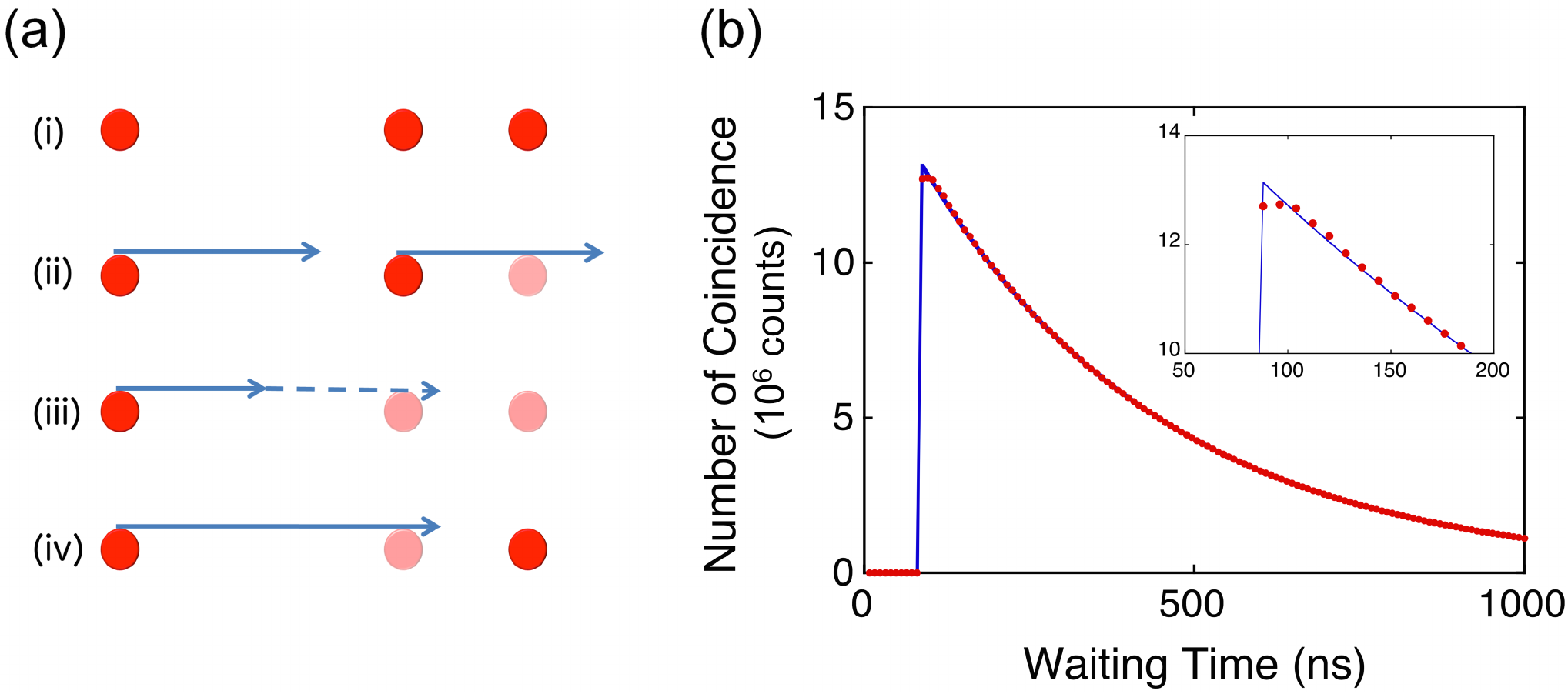}
\caption{
Over-deletion problem and its effect on the waiting-time distribution.
In (a), the red -filled circles represent photodetection events in sequence whereas the pink-filled circles indicate missed events due to an intrinsic or a prolonged dead time.
(i). Dead-time-free case. (ii). With an intrinsic dead time, whose length is indicated by a solid arrow. The third event is missed due to the intrinsic dead time. (iii). With a prolong dead time. The prolonged part from the intrinsic dead time is marked by a dashed arrow. The second event within the prolonged dead time is deleted. (iv). With an intrinsic dead time as long as the prolong dead time in (iii). Differently from case (iii), the third event is detected. Therefore, over-deletion occurs in (iii) compared to the case of a real dead time.
(b) Waiting time distribution with a 80-ns prolonged dead time from a 28-ns intrinsic dead time. A small distortion, marked by a blue arrow, occurs around 80 ns, but it is so small that the waiting time distribution can still be approximated by a single exponential function.
}
\label{fig5}
\end{figure}

\subsection{Simulating Prolonged Detector Dead Times}

In order to investigate the detector dead time effect systematically in experiment, it is desired to have as many detectors with the same characteristics but with different dead times as possible. In reality, we have a limited number of detectors. We have two SPCM-F's  with a mean dead time of 28-ns and two SPCM-S's with a 56-ns dead time.

To overcome this limitation, we have simulated additional dead times for a given detector by deleting the subsequent photodetection records within an extended period beyond the actual dead time after any photodetection event. Those extended periods serve as prolonged dead times.

However, the effect of a prolonged dead time is not exactly the same as that of a real dead time with an equal magnitude. 
It is because with a prolong dead time some photodetection events are lost which would be detected with a real dead time with the same magnitude as illustrated in Fig.\ \ref{fig5}(a). This over-deletion of counts leads to a distortion in the waiting time distribution function near the prolonged dead time as indicated in Fig.\ \ref{fig5}(b). Nonetheless, the distortion is not large enough to affect the overall shape of the waiting time distribution, indicating the over-deletion rarely occurs.
Therefore, we can utilize the prolonged dead times for systematic investigation of the dead time effect in the next section.

\begin{figure}
\includegraphics[width=3.4in]{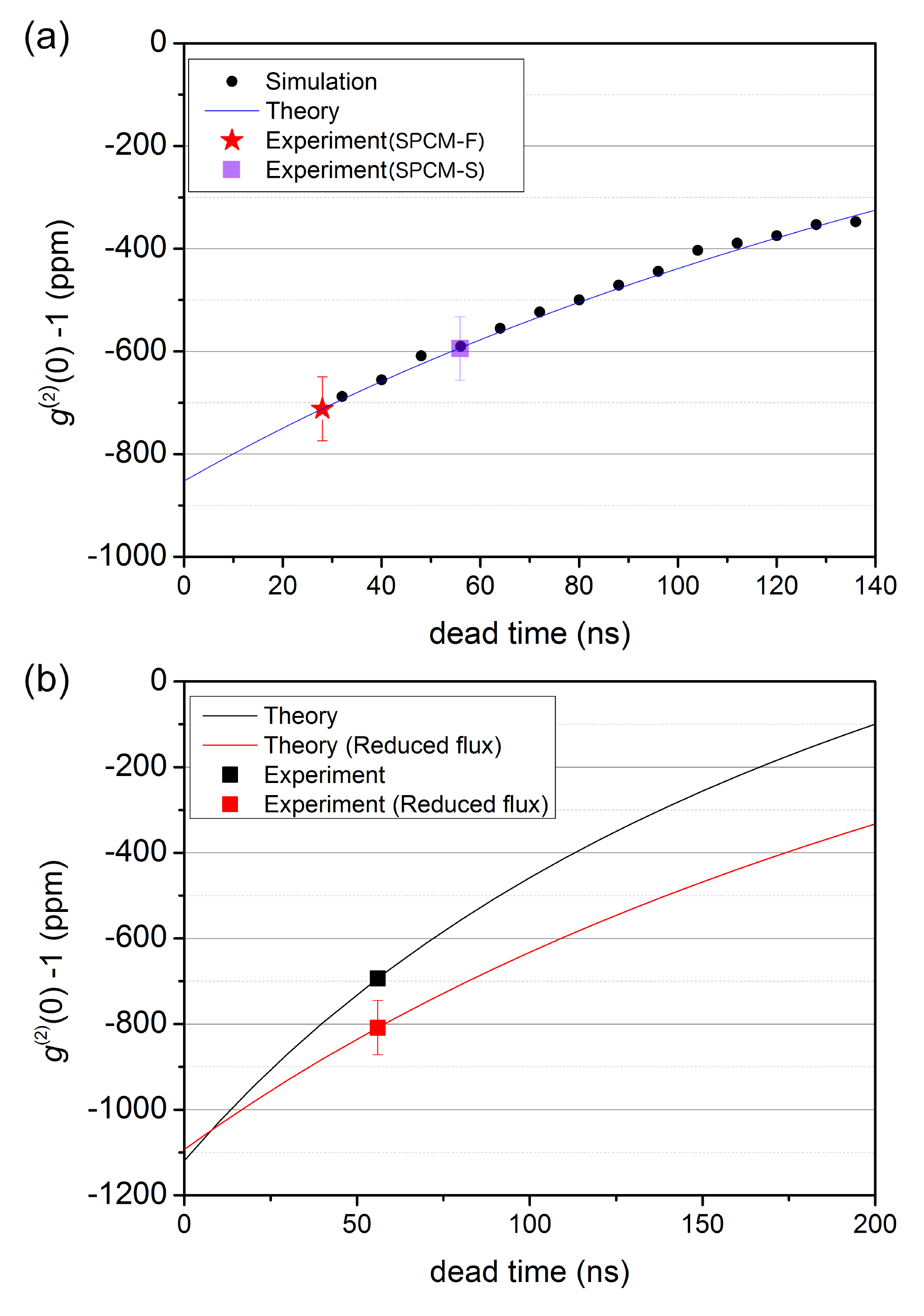}
\caption{
Dead time effect on the SOC function of the cavity-QED microlaser.
(a) Dependence of the observed SOC on intrinsic and prolonged dead times.
Red star is the observed $g'^{(2)}(0)$ with SPCM-F's of a 28-ns dead time whereas the purple square is that with SPCM-S's of a 56-ns dead time.
Black dots are the results obtained with prolonged dead times applied to the photodetection records corresponding to the data marked by the red star. 
Blue line is a theoretical fit given by Eq.\ (\ref{eq6}) under the condition that it should go through the red dot.
(b) Dependence of the observed SOC on photodetection flux.
Black(red) square is the observed $g'^{(2)}(0)$ with SPCM-S's when the photo fluxes on start and stop detectors are 2.75(1.94) Mcps and 2.23(1.60) Mcps, respectively.
Black(red) solid curve is a theoretical fit given by Eq.\ (\ref{eq6}) under the condition that it should go through the black(red) square.
Error bars represent the fitting errors in obtaining $g'^{(2)}(0)$ values from the SOC data. 
Typical fitting errors for simulated data with prolonged dead times are normally 10\% of $\left | g^{(2)}(0)-1 \right |$. 
The error bar for the black square in (b) is smaller than than size of the square.
}
\label{fig6}
\end{figure}

\section{Results and Discussions}\label{sec5}

We have measured $g'^{(2)}(0)$ (in the presence of the dead time effect) of the cavity-QED microlaser output by using the setup depicted in Fig.\ \ref{fig4} with two different sets of detectors. The results are shown in Fig.\ \ref{fig6}(a), where the red star indicates the result with SPCM-F's  and the purple square corresponds to the result with SPCM-S's. The dead-time-corrected photodetection flux on each detector was 2.6 Mcps for SPCM-F and 3.3 Mcps for SPCM-S.  Also plotted in Fig.\ \ref{fig6}(a) as black dots are the results obtained with prolong dead times as discussed in Sec.\ \ref{sec4}.C on SPCM-F's of 28-ns dead time. A solid curve is a theoretical fit by Eq.\ (\ref{eq6}), well agreeing with the results with actual and prolong dead times.  
The only fitting parameter is $g^{(0)}(0)$, which appears as a vertical axis offset corresponding to zero dead time. 
The dead-time-free fluxes needed in Eq.\ (\ref{eq6}) are obtained from the observed photodetection fluxes by using Eq.~(\ref{eq1}).

The smallest $g'^{(2)}(0)-1$ measured with SPCM-F is 
-710$\pm$60 ppm (part per million), corresponding to Mandel $Q$ of 
-0.43$\pm$0.04, whereas the dead-time-corrected $g^{(2)}(0)-1$ obtained from the fitting is 
-850$\pm$60 ppm and thus the actual Mandel $Q$ of the microlaser output is 
-0.51$\pm$0.04. It is noteworthy that the smallest dead time effect with our best detector still amounts to a considerable distortion (0.51/0.43-1=19\%) in the Mandel $Q$ measurement. The dead time correction by Eq.\ (\ref{eq6}) is thus essential for accurate photon statistics measurement.  

In Fig.\ \ref{fig6}(b), we examine the dependence of $g'^{(2)}(0)$ on photodetection flux.
The black and red squares represent  the results of $g^{(2)}(0)$ measurement with the same detectors (SPCM-S) but with a reduce flux for the red one. The solid curves are theoretical fits by Eq.\ (\ref{eq6}) with $g^{(2)}(0)$ as a fitting parameter individually. The fit curves, although independently done, almost meet at zero dead time. The discrepancy is less than the experimental error.

The observed dependence $g'^{(2)}(0)$ on the photodetection flux may appear contradictory to the general view that the SOC function does not depend on random miss of incident photons. An example is detector efficiency.
It should be noted, however, that the missing of incident photons due to the detector dead time is not random miss at all. The missing occurs only immediately after a successful photodetection event. In other words, the missing event is correlated with the success event with the detector dead time as a correlation time. So, there is no contradiction.

One way to avoid the dead time effects discussed so far is to keep the photodetection flux low enough to make the mean waiting time very much larger than the dead time, $\tau_w\ggg \tau_d$. The distortion in SOC will then be negligible as shown in Eq.~(\ref{eq7}). 
This approach does not work, however, when the measurement time $T_0$ for SOC is practically limited. 
For instance, a photon source may have a finite operating time and thus $T_0$ is limited.
In order to resolve the feature in $g^{(2)}(t)$ near the origin for Mandel $Q$ measurement, the signal-to-noise ratio has to be larger than $|1- g^{(2)}(0)|^{-1}$.
For a bin time $t_b\ll \tau_c$, we then have to satisfy 
$\sqrt{(T_0/\tau_w)^2/(T_0/t_b)} |1- g^{(2)}(0)|>1$ or $\tau_w <\sqrt{T_0 t_b} |1- g^{(2)}(0)|$.
This requirement set a upper bound for $\tau_w$. 
If this upper bound is not much larger than the detector dead time, we cannot avoid the dead time effect and thus have to rely on our correction formula.
In the original HBT experiment, the photon flux from a distant star were extremely low. Consequently, the waiting time $\tau_w$ was much larger than the detector dead time and thus its effect was negligibly small.

\section{Conclusion}\label{sec6}

We have found that detector dead time can induce a significant error in the measurement of $g^{(2)}(t)$ of a stationary light source even in two detectors configuration when the internal mean photon number is much larger than unity.
This finding contradicts the conventional understanding that detector dead time would not affect $g^{(2)}(t)$ in two-detector configurations.
In experiment, we employed the cavity-QED microlaser for a sub-Poissonian light source and measured $g^{(2)}(0)$ with two different types of photodetectors with different dead time. 
The observed $Q$ of the cavity-QED microlaser was underestimated as much as 19\% with respect to the dead-time-free $Q$ even when we used single-photon counting modules with the shortest dead time available. 
We also simulated prolonged dead times by intentionally deleting the photodetection events following a preceding one.
The observed values of $g^{(2)}(0)$ for various real and prolonged dead time were well explained by 
our analytic formula.
Dead-time-free $g^{(2)}(0)$ and thus Mandel $Q$ of a stationary light source can be obtained with our correction formula.

The present work is limited to the case of a large internal mean photon number so as to neglect the higher-order correlation. 
By including the higher-order correlation, one may obtain a more general formula for the detector dead time effect on $g^{(2)}(0)$.\\

\hbox{}

\appendix* \label{app}
\section{Conditions for approximating the waiting time distribution as a single exponential}

Let us consider the Laplace transforms $W(s)$ and $G(s)$ of a waiting-time distribution $w(t)$ and a SOC function $g^{(2)}(t)$, respectively. 
According to  Ref.~\cite{Ververk-JCP03}, they are related as
\begin{equation}
W(s)=\phi G(s)/(1+\phi G(s)). 
\label{eq8}
\end{equation}
For the cavity-QED microlaser operating at a high mean photon number, $g^{(2)}(t)$ can be written as
\begin{equation}
g^{(2)}(t)=1-\beta e^{-t/\tau_c}.
\label{eq9}
\end{equation}
Then $G(s)$ is readily given by 
\begin{equation}
G(s)=\frac{1}{s} + \frac{1-\beta}{s+a},
\label{eq10}
\end{equation}
where $a=1/\tau_c$. 
Inserting this expression in Eq.~\ (\ref{eq8}) and taking inverse Laplace transform, we obtain
the waiting-time distribution of the microlaser as
\begin{equation}
w(t)=\phi \mathcal{L}^{-1}\left[ \frac{(1-\beta)s+ a }{s^2+\left[a+\phi(1-\beta)\right]s+\phi a} \right].
\label{eq11}
\end{equation}
The above waiting-time distribution can be approximated by a single exponential $\phi e^{-\phi t}$ if $\beta \ll 1$, which can be expressed as $|1-g^{(2)}(0)|\ll 1$. \\


\end{document}